\documentclass[english,pre,twocolumn,nofootinbib,superscriptaddress]{revtex4-1}
\usepackage[T1]{fontenc}
\usepackage[latin9]{inputenc}
\usepackage{geometry}
\usepackage{color}
\usepackage{babel}
\usepackage{amsmath}
\usepackage{amssymb}
\usepackage{graphicx}
\usepackage[unicode=true,pdfusetitle,
 bookmarks=true,bookmarksnumbered=false,bookmarksopen=false,
 breaklinks=true,pdfborder={0 0 0},pdfborderstyle={},backref=false,colorlinks=true]
 {hyperref}
\begin{document}

\title{Dynamical Mean Field Theory, Density-Matrix Embedding Theory and
Rotationally Invariant Slave Bosons: a Unified Perspective}

\author{Thomas Ayral}
\affiliation{Physics and Astronomy Department, Rutgers University, Piscataway, New Jersey 08854, USA}

\author{Tsung-Han Lee}
\affiliation{Physics and Astronomy Department, Rutgers University, Piscataway, New Jersey 08854, USA}

\author{Gabriel Kotliar}
\affiliation{Physics and Astronomy Department, Rutgers University, Piscataway, New Jersey 08854, USA}
\affiliation{Condensed Matter Physics and Materials Science Department, Brookhaven National Laboratory, Upton, New York 11973, USA}

\begin{abstract}
We present a unified perspective on Dynamical Mean Field Theory (DMFT),
Density-Matrix Embedding Theory (DMET) and Rotationally Invariant
Slave Bosons (RISB). We show that DMET can be regarded as a simplification
of the RISB method where the quasiparticle weight is set to unity. 
This relation allows to easily transpose extensions of a given method to another: for instance, a temperature-dependent version of RISB can be used to derive a temperature-dependent free-energy formula for DMET.
\end{abstract}

\maketitle
Strong correlations count amongst the most challenging problems in condensed-matter
physics. While the development of Dynamical Mean Field Theory (DMFT) \cite{Georges1996}
and its cluster \cite{Hettler1998,Hettler1999,Lichtenstein2000,Kotliar2001,Maier2005a}
and diagrammatic \cite{Rohringer2017} extensions has led to a better
understanding of relatively simple strongly-correlated models and
systems, there are still situations where the exact solution of the
DMFT quantum impurity model becomes prohibitive due to the size of
its Hilbert space and/or the Monte-Carlo negative sign problem. These
situations range from the study of multiorbital systems to the exploration
of low temperature phases over the investigation of long-range strong
correlations. This is particularly important for realistic investigations
of 5f systems, which require the  simultaneous  inclusion of crystal-field effects,
spin-orbit coupling interaction multiplets and lattice relaxation.
Outstanding challenges in this area include the computation of phase diagrams and equations of state  of elemental actinides and their
alloys.   These are problems of fundamental importance and of practical technological relevance, and that require simplified  faster  methods  that still 
capture correlation effects accurately enough. 

Several such methods have been developed in  recent years, with
a commonality with DMFT: the mapping of the lattice problem onto a
simpler, yet still nontrivial embedded   quantum problem. Prominent examples
include cluster perturbation theory (CPT, \cite{Gros1993,Senechal2000},
derivable from the self-energy functional theory, SFT \cite{Potthoff2003,Potthoff2012}),
self-energy embedding theory (SEET, \cite{Zgid2016,Lan2017}), two-site
DMFT \cite{Potthoff2001,Moeller1994}, and site-occupation embedding
theory (SOET, \cite{Fromager2015,Senjean2016}). 

Two particularly successful methods are the (mean-field) rotationally
invariant slave-boson method (RISB, \cite{Fresard1992,Lechermann2007}) and the
density-matrix embedding theory (DMET, \cite{Knizia2012,Knizia2013}).
RISB yields  kinetic energy renormalizations,  double occupancies and valence histograms
very close to DMFT \cite{Ferrero2008,Ferrero2009,Mazin2014}
and has been applied to numerous multiband models \cite{Lechermann2007,Lanata2012,Facio2017,Piefke2017}
and realistic compounds \cite{Lanata2015,Lanata2016,Piefke2017}. 
These slave-boson methods have a close connection to the Gutzwiller approximation
as shown in Ref.~\cite{Kotliar1986} for the single-site case and in Ref.~\cite{Bunemann2007} for the  multiorbital case.
DMET has been shown to yield very accurate ground-state energies
for the Hubbard model \cite{Knizia2012,Knizia2013,Zheng2016,Zheng2017,Bulik2014}
and quantum chemical systems \cite{Wouters2016a,Wouters2016}. Both
allow to reach ground-state (including superconducting \cite{Isidori2009,Mazza2017,Zheng2016})
properties and spectral properties~\cite{Booth2015}, and have also
been extended to tackle out-of-equilibrium problems \cite{Schiro2010,Schiro2011,Lanata2012a,Mazza2012,Behrmann2013,Mazza2015,Behrmann2015,Kretchmer2016}.

However, the precise relation between these two methods has not been established
to date and it is unclear whether they yield a complementary picture
of correlations, or if on the contrary one corresponds to the simplification
of the other. This work intends to fill this gap by showing that DMET
is a simplication of RISB. We also illustrate the relation of RISB
with DMFT, thereby giving a comprehensive picture of the interrelation
and hierarchy between the three methods.

This paper is organized as follows: we first give an overview of the
results presented in this paper (section \ref{sec:Overview}), then
review the RISB formalism and its relation with DMFT (section \ref{sec:Overview-of-rotationally}),
and finally derive the DMET approximation and show that it is a simplified
RISB with a quasiparticle weight equal to one (section \ref{sec:Density-matrix-embedding-theory:}). 

\section{Overview\label{sec:Overview}}

In this section, we highlight the common structure of RISB and DMET
without providing detailed derivations. Our purpose is to provide
the reader with the key ideas that these methods share and thus reveal
their close connection.

\begin{figure}
\begin{centering}
\includegraphics[width=1\columnwidth]{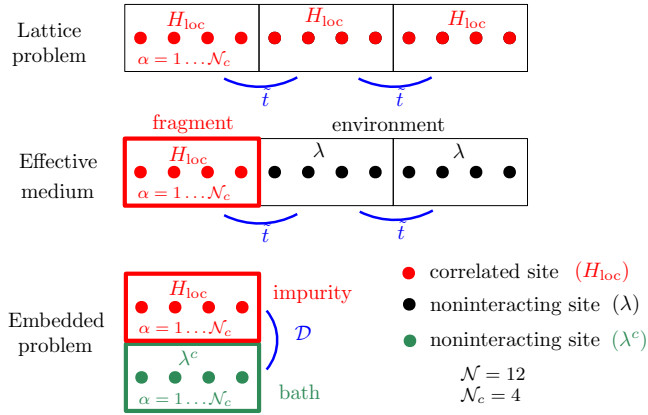}
\par\end{centering}
\caption{Three layers of RISB and DMET. \emph{Top}: Lattice problem: all unit
cells are interacting. \emph{Middle}: Effective medium: only the unit
cell dubbed ``fragment'' is interacting, correlations for the rest
(``environment'') are described by a one-body potential $\lambda$.
\emph{Bottom}: Embedded problem: the environment is mapped to a ``bath''
of the same dimension as the fragment/impurity.\label{fig:DMET-fragment-environment}}
\end{figure}

Both RISB and DMET start with an \emph{interacting lattice model}
of the form

\begin{equation}
\hat{H}=\sum_{ij,\alpha\beta}\tilde{t}_{i\alpha,j\beta}c_{i\alpha}^{\dagger}c_{j\beta}+\sum_{i}\hat{H}_{\mathrm{loc}}[\{c_{i\alpha},c_{i\alpha}^{\dagger}\}],\label{eq:Hamiltonian-with-H-loc}
\end{equation}
and depicted in Fig. \ref{fig:DMET-fragment-environment}, top panel. Greek indices $\alpha,\beta,\dots=1\dots \mathcal{N}_c$ denote local or orbital indices (within a unit cell). Latin indices $i,j,\dots \mathcal{N}/\mathcal{N}_c$ denote unit cells. (The local or orbital degrees of freedom may as well refer to inequivalent sites in a cellular-DMFT-like construction, so that a "unit cell" may also refer to a cluster cell).
The first term is a kinetic term describing hopping processes between different unit cells. The hopping $\tilde{t}$ does not contain local hopping terms ($\tilde{t}_{i\alpha,i\beta}=0$); instead, they are contained in $\hat{H}_\mathrm{loc}$. Later, we will denote by $t$ the full hopping matrix, and by $\hat{H}_\mathrm{int}$ the interaction Hamiltonian (without local hoppings). 

The key idea of RISB and DMET is to replace the lattice model by a
reference model or \emph{effective medium} that depicts correlations
in an approximate fashion.

In DMET, this effective medium consists of one correlated unit cell,
called the \emph{fragment}, and an \emph{environment} where the effect
of correlations is described by a one-body potential $\lambda$, as illustrated in Fig. \ref{fig:DMET-fragment-environment},
middle panel. (In the DMET literature, this potential is usually called
$u$ and is equal to $\lambda$ up to a local hopping term: $\lambda_{\alpha\beta}=u_{\alpha\beta}+\left[\varepsilon_{\mathrm{loc}}\right]_{\alpha\beta}$.) The Hamiltonian of this effective medium is:

\begin{align}
 & \hat{H}_{\mathrm{eff}}\equiv\label{eq:H_DMET}\\
 & \sum_{ij,\alpha\beta}\tilde{t}_{i\alpha,j\beta}c_{i\alpha}^{\dagger}c_{j\beta}+\hat{H}_{\mathrm{loc}}[\{c_{0\alpha},c_{0\alpha}^{\dagger}\}]+\sum_{i\neq0,\alpha\beta}\lambda_{\alpha\beta}c_{i\alpha}^{\dagger}c_{i\beta}\nonumber .
\end{align}

RISB also has an effective medium, but it comes with an additional
parameter (called $R$ in the following) that allows to describe the
mass of quasiparticles and cannot be described by a one-body potential
only.

As a last step, both methods introduce an impurity or \emph{embedded
problem} illustrated in Fig. \ref{fig:DMET-fragment-environment},
bottom panel. It is obtained by the contraction of the environment (of size $\mathcal{N}-\mathcal{N}_c$)
to $\mathcal{N}_c$ \emph{bath} orbitals using a Schmidt decomposition. The embedded problem thus consists of $\mathcal{N}_c$
correlated or \emph{impurity} orbitals hybridized (via a hybridization term
$\mathcal{D}$) to $\mathcal{N}_c$ uncorrelated bath orbitals (described by the one-body
potential $\lambda^{c}$). It is given by a Hamiltonian of the form

\begin{align}
 & \hat{H}_{\mathrm{embed}}\equiv\label{eq:H_emb_def}\\
 & \;\;\;\sum_{\alpha\beta}\left(\mathcal{D}_{\alpha\beta}c_{\alpha}^{\dagger}a_{\beta}+\mathrm{h.c}\right)+\hat{H}_{\mathrm{loc}}[\{c_{\alpha}^{\dagger},c_{\alpha}\}]+\sum_{\alpha\beta}\lambda_{\alpha\beta}^{c}a_{\beta}a_{\alpha}^{\dagger}\nonumber ,
\end{align}

\noindent Here, $c^{\dagger}$/$c$ (resp. $a^{\dagger}$/$a$) are creation/annihilation
operators for the impurity (resp. bath) orbitals.

The goal of both methods is to determine the effective medium, i.e
to find the value of $\lambda$ (and optionally $R$), such that the
following self-consistency condition is satisfied: the one-particle \emph{density matrix}
of the impurity model (Eq. (\ref{eq:H_emb_def})) must match the projection
of the reference medium's density matrix onto the embedded subspace.
$\lambda$ (and optionally $R$) can then be used as approximations
of the self-energy of the lattice model.

This is also the logic of DMFT, except that DMFT adjusts the local
self-energy $\Sigma_{\mathrm{loc}}(i\omega)$ (instead of $\lambda$
and $R$ above) so that the (one-particle) \emph{Green's function},
\begin{equation}
\left[G_{\mathrm{imp}}\right]_{\alpha\beta}(i\omega_{n})\equiv-\int_{0}^{\beta}\mathrm{d}\tau e^{i\omega_{n}\tau}\langle Tc_{\alpha}(\tau)c_{\beta}^{\dagger}(0)\rangle_{\mathrm{imp}}\label{eq:G_imp_def},
\end{equation}
of the impurity model matches the projection of the reference medium's
Green's function,
\begin{equation}
G(\boldsymbol{k},i\omega)=\left[i\omega-\varepsilon_{\boldsymbol{k}}-\Sigma_{\mathrm{loc}}(i\omega)\right]^{-1}\label{eq:G_latt_DMFT},
\end{equation}
onto the impurity (where $\varepsilon_{\boldsymbol{k}}$ is the Fourier
transform of the hopping matrix $t_{ij}$ or LDA Hamiltonian in a LDA+DMFT context, see Eq.~\eqref{eq:FT_def} below).
Besides, because of this modified self-consistency condition, the impurity model of DMFT contains an infinite number of
bath sites, contrary to Eq. (\ref{eq:H_emb_def}).

While connections between RISB, DMET and DMFT were mentioned in passing
(see e.g. \cite{Lanata2015}) and DMET was inspired by DMFT, a precise
direct connection was not available in the literature to this date.
This connection points to many possible generalizations of these methods.

\section{Overview of rotationally invariant slave bosons\label{sec:Overview-of-rotationally}}

In this section, we briefly review the RISB formalism introduced by
\cite{Lechermann2007}. Our starting point is the lattice Hamiltonian
Eq. (\ref{eq:Hamiltonian-with-H-loc}). We note that $\hat{H}_{\mathrm{loc}}$
contains both kinetic and interaction terms. Denoting by $t_{i\alpha,i\beta}$
hoppings internal to a unit cell, we can decompose: 

\begin{align}
\hat{H}_{\mathrm{loc}} & =\hat{H}_{\mathrm{int}}+\sum_{i}\sum_{\alpha\beta}t_{i\alpha,i\beta}c_{i\alpha}^\dagger c_{i\beta}\label{eq:H_loc_def},\\
\tilde{t}_{i\alpha,j\beta} & =\begin{cases}
t_{i\alpha,j\beta} & \mathrm{if\;}i\neq j\\
0 & \mathrm{otherwise}
\end{cases}\nonumber .
\end{align}

\noindent In the following, we will denote the $i,j$ Fourier transform of $t_{i\alpha,j\beta}$
(resp. $\tilde{t}_{i\alpha,j\beta}$) as 
\begin{equation}
\varepsilon_{\boldsymbol{k},\alpha\beta}\equiv\frac{\mathcal{N}_c}{\mathcal{N}}\sum_{ij}e^{-i\boldsymbol{k}\cdot(\boldsymbol{r}_{i}-\boldsymbol{r}_{j})}t_{i\alpha,j\beta}\label{eq:FT_def}
\end{equation}
(resp. $\tilde{\varepsilon}_{\boldsymbol{k},\alpha\beta}$). Correspondingly,
\begin{equation}
\left[\varepsilon_{\mathrm{loc}}\right]_{\alpha\beta}\equiv\sum_{\boldsymbol{k}}\varepsilon_{\boldsymbol{k},\alpha\beta}=t_{0\alpha,0\beta}\label{eq:eps_loc_def}.
\end{equation}

\subsubsection{Slave bosons: Constraints and physical subspace}

The second-quantized operators $c_{i\alpha}^{\dagger}$, $c_{i\alpha}$
generate a Hilbert space $\mathcal{H}_{\mathrm{phys}}$ with \emph{local}
many body states $|A_{i}\rangle$. 
RISB consists in introducing fermionic operators $f_{i\alpha}^{\dagger}$,
$f_{i\alpha}$ and bosonic operators $\hat{\phi}_{A_{i},n_{i}}^{\dagger}$,
$\hat{\phi}_{A_{i},n_{i}}$ (one for each \emph{pair} of \emph{local} many-body states, with $n_i$ labelling local Fock states, which like $A_i$ form a basis of the local Hilbert space) to replace $c$ and $c^\dagger$. Yet, these new operators generate a Hilbert space $\mathcal{H}$ which is much
larger than the original Hilbert space, so that one needs to define
a ``physical'' subspace. This is done by defining \emph{physical
states} and the corresponding \emph{constraints}. The physical states
are defined as follows:
\begin{equation}
|\underline{A}_{i}\rangle\equiv\frac{1}{\sqrt{D_{A}}}\sum_{n_{i}}\hat{\phi}_{A_{i},n_{i}}^{\dagger}|0\rangle|n_{i}\rangle_{f}\label{eq:physical_states_def},
\end{equation}
with $D_{A}\equiv\left(\begin{array}{c} 
\mathcal{N}_c\\
\mathcal{N}_A
\end{array}\right)$ ($\mathcal{N}_A$ is the number of electrons in state $A$) and the $\lbrace|n_i\rangle_f\rbrace$ are the local Fock states formed with $f^\dagger$ operators:
\begin{equation}
|n_{i}\rangle_f\equiv\prod_{\alpha=1}^{\mathcal{N}_c}(f_{i\alpha}^{\dagger})^{n_{i\alpha}}|0\rangle.
\end{equation}One can check the physical states $|\underline{A}_i\rangle$ are normalized. One can prove~\cite{Lechermann2007}
that these states (and only these states) satisfy the following constraints:\begin{subequations}

\begin{align}
\forall i\;\;\; & \sum_{A_{i}n_{i}}\hat{\phi}_{A_{i}n_{i}}^{\dagger}\hat{\phi}_{A_{i}n_{i}}=1\label{eq:constraint_1},\\
\forall i\;\;\; & f_{i\alpha}^{\dagger}f_{i\beta}=\sum_{A_{i}n_{i}m_{i}}{}_f\langle m_{i}|f_{i\alpha}^{\dagger}f_{i\beta}|n_{i}\rangle_f\hat{\phi}_{A_{i}n_{i}}^{\dagger}\hat{\phi}_{A_{i}m_{i}}.\label{eq:constraint_2}
\end{align}
\end{subequations}

\noindent In the following, we drop the site index $i$ for conciseness.

One then writes a \emph{faithful representation} of the original Hamiltonian
$H$ in the physical subspace, where ``faithful'' is defined as
follows: for any (local) operator $O$, $\underline{O}$ is said to
be a faithful representation of $O$ in the physical subspace if and
only if:
\begin{equation}
\forall A,B,\;\;\;\langle\underline{A}|\underline{O}|\underline{B}\rangle=\langle A|O|B\rangle\label{eq:faithful_repr_RISB}.
\end{equation}

\noindent One can show~\cite{Lechermann2007} that the faithful representation
of the creation operator is given by the expression:

\begin{equation}
\underline{c}_{\alpha}^{\dagger}=\sum_{\beta}R_{\beta\alpha}f_{\beta}^{\dagger}\label{eq:faithful_RISB-1},
\end{equation}
with the $\hat{\phi}$-dependent matrix $R$ defined by:
\begin{equation}
R_{\beta\alpha}\equiv\sum_{ABnm}\frac{\langle A|c_{\alpha}^{\dagger}|B\rangle{}_f\langle n|f_{\beta}^{\dagger}|m\rangle_f}{\sqrt{\mathcal{N}_A(\mathcal{N}_c-\mathcal{N}_B)}}\hat{\phi}_{An}^{\dagger}\hat{\phi}_{Bm}\label{eq:R_def}.
\end{equation}

\noindent Eq. (\ref{eq:faithful_RISB-1}) is used to write down the faithful
representation of the kinetic term of the Hamiltonian. For the local Hamiltonian
term (which contains interactions and local hoppings, see above), one can show \cite{Lechermann2007} that 
\begin{equation}
\underline{H}_{\mathrm{loc}}=\sum_{iABn}\langle A_{i}|H_{\mathrm{loc}}|B_{i}\rangle\hat{\phi}_{A_{i}n_{i}}^{\dagger}\hat{\phi}_{B_{i}n_{i}}\label{eq:faithful_h_int_RISB}.
\end{equation}

\noindent One can thus write the faithful representation of $H$ in terms of
the $f$ and $\hat{\phi}$ fields:
\begin{align}
\underline{H} & =\sum_{ij,\gamma\delta}\left\{ \sum_{\alpha\beta}R_{\gamma\alpha}\tilde{t}_{i\alpha,j\beta}R_{\beta\delta}^{\dagger}\right\} f_{i\gamma}^{\dagger}f_{j\delta}\nonumber \\
 & \;\;+\sum_{i,AB}\langle A_{i}|H_{\mathrm{loc}}|B_{i}\rangle\sum_{n_{i}}\hat{\phi}_{A_{i}n_{i}}^{\dagger}\hat{\phi}_{B_{i}n_{i}}\label{eq:H_faithful_RISB}.
\end{align}

\noindent This Hamiltonian is nontrivial in the $\hat{\phi}$ operators (through the
$\hat{\phi}$-dependence of $R$), but quadratic in the $f$ operators.
In the following, we will thus carry out a mean-field approximation
for the bosons, and integrate out the $f$ fields.

\subsubsection{Mean-field approximation and matrix notation}

We now condense the bosons, i.e $\hat{\phi}_{Bn}$ is chosen to be a $c$-number
(and $\hat{\phi}_{Bn}^{\dagger}$ becomes $\phi_{Bn}^{*}$). For notational
convenience, we define the $2^{\mathcal{N}_c}\times2^{\mathcal{N}_c}$ matrices $\Phi$ and
$F$:
\begin{align}
[\Phi]_{An} & \equiv\phi_{An}\label{eq:Phi_matrix_def-1},\\{}
[F_{\alpha}]_{nm} & \equiv{}_{f}\langle n|f_{\alpha}|m\rangle_{f}.\label{eq:F_matrix_def-1}
\end{align}

\noindent In particular, $[\Phi^{\dagger}]_{nA}=[\Phi]_{An}^{*}=\phi_{An}^{*}$.
We can always order the $|A\rangle$ states in such a way that $\langle A|c_{\alpha}|B\rangle=[F_{\alpha}]_{AB}$.
In particular, $[F_{\alpha}^{\dagger}]_{AB}=[F_{\alpha}]_{BA}^{*}=(\langle B|c_{\alpha}|A\rangle)^{*}=\langle A|c_{\alpha}^{\dagger}|B\rangle$.
If the coefficients of the $F$ matrix are real (which is the case
if one is dealing with Fock states, which is always possible), then
\begin{equation}
[F_{\alpha}^{\dagger}]_{AB}=[F_{\alpha}]_{BA}\label{eq:F_real-1}.
\end{equation}

\noindent Thus, the expressions for the constraints become:\begin{subequations}

\begin{align}
\mathrm{Tr}[\Phi\Phi^{\dagger}] & =1\label{eq:constraint_1-1},\\
f_{\alpha}^{\dagger}f_{\beta} & =\Delta_{\alpha,\beta}^{p}\;\;\forall\alpha,\beta.\label{eq:constraint_2-1}
\end{align}
\end{subequations}with

\begin{align}
\Delta_{\alpha\beta}^{p} & \equiv\sum_{Anmp}{}_f\langle m|f_{\alpha}^{\dagger}|p\rangle_f{}_f\langle p|f_{\beta}|n\rangle_f[\Phi^{\dagger}]_{nA}\Phi_{Am}\nonumber \\
 & =\mathrm{Tr}\left[F_{\alpha}^{\dagger}F_{\beta}\Phi^{\dagger}\Phi\right].\label{eq:Delta_p_constraint}
\end{align}

\noindent Furthermore,
\begin{align*}
R_{\beta\alpha} & =\sum_{\gamma}\sum_{ABnm}F_{\alpha,A,B}^{\dagger}F_{\gamma,nm}^{\dagger}[\Phi^{\dagger}]_{nA}[\Phi]_{Bm}\\
 & \;\;\;\;\times\left[\left(\Delta^{p}(1-\Delta^{p})\right)^{-1/2}\right]_{\gamma\beta}\\
 & =\sum_{\gamma}\mathrm{Tr}\left[\Phi^{\dagger}F_{\alpha}^{\dagger}\Phi F_{\gamma}\right]\left[\left(\Delta^{p}(1-\Delta^{p})\right)^{-1/2}\right]_{\gamma\beta}.
\end{align*}

\noindent where, to obtain the second line, we have assumed $F$ to be real valued
(Eq. (\ref{eq:F_real-1})). Equivalently, we have:

\begin{equation}
R_{\gamma\alpha}\left[\left(\Delta^{p}(1-\Delta^{p})\right)^{1/2}\right]_{\gamma\beta}=\mathrm{Tr}\left[\Phi^{\dagger}F_{\alpha}^{\dagger}\Phi F_{\beta}\right].\label{eq:R_multiplied_constraint}
\end{equation}

\noindent The local part of the Hamiltonian reads:
\begin{align}
\underline{H}_{\mathrm{loc}} & =\sum_{i}\sum_{A_{i}B_{i}n_{i}}\langle A_{i}|H_{\mathrm{loc}}|B_{i}\rangle[\Phi^{\dagger}]_{n_{i}A_{i}}\Phi_{B_{i}n_{i}}\nonumber \\
 & =\sum_{i}\mathrm{Tr}[\Phi^{\dagger}H_{\mathrm{loc}}\Phi]\label{eq:H_loc_as_trace}
.\end{align}

\subsubsection{Mean field free-energy and Lagrange equations}

The problem at hand now boils down to minimizing the free energy,
a function of the slave-boson mean fields $\Phi_{An}$, under the
constraints (\ref{eq:constraint_1-1}-\ref{eq:constraint_2-1}). In
the original formulation~\cite{Lechermann2007}, inspired by previous
slave-boson approaches~\cite{Kotliar1986}, the fulfillment of the
constraints was enforced by introducing two Lagrange multipliers $E^{c}$
and $\lambda$. It was then proposed \cite{Lanata2012,Lanata2015,Lanata2016},
in order to overcome the remaining strong nonlinearity of the free
energy as a function of $\Phi$, to turn Eqs (\ref{eq:Delta_p_constraint}-\ref{eq:R_multiplied_constraint})
into constraints, thereby making the free energy quadratic in $\Phi$ at the price of adding two more Lagrange multipliers $\lambda^{c}$
and $\mathcal{D}$ and turning $\Delta^{p}$ and $R$ into independent
variables. Following
this strategy, the free energy of the system is given by:
\begin{align}
 & \Omega[\Phi,R,\Delta^{p};E^{c},\lambda,\mathcal{D},\lambda^{c}]\equiv\nonumber \\
 & \;\;\;-\beta\log\int\mathcal{D}[f^{*},f]e^{-S[\Phi,R,\Delta^{p};E^{c},\lambda,\mathcal{D},\lambda^{c}]}\label{eq:free_energy},
\end{align}

\noindent with 
\begin{align}
 & S[\Phi,R,\Delta^{p};E^{c},\lambda,\mathcal{D},\lambda^{c}]=\nonumber \\
 & -\sum_{\boldsymbol{k}i\omega}\mathrm{Tr}\log\left\{ i\omega-R_{\alpha\gamma}\tilde{\varepsilon}_{\boldsymbol{k}}^{\gamma\delta}R_{\delta\beta}^{\dagger}-\lambda_{\alpha\beta}+\mu\delta_{\alpha,\beta}\right\} e^{i\omega0^{+}}\nonumber \\
 & +\sum_i\mathrm{Tr}\Bigg[E^{c}\left(\Phi^{\dagger}\Phi-1\right)+\sum_{\alpha\beta}\left(\mathcal{D}_{\alpha\beta}\Phi^{\dagger}F_{\alpha}^{\dagger}\Phi F_{\beta}+\mathrm{h.c}\right)\nonumber \\
 & \;\;\;\;\;\;\;\;\;\;+\sum_{\alpha\beta}\lambda_{\alpha\beta}^{c}\Phi^{\dagger}\Phi F_{\alpha}^{\dagger}F_{\beta}+\Phi^{\dagger}H_{\mathrm{loc}}\Phi\Bigg]\nonumber \\
 & -\sum_{i;\alpha\beta}\left(\lambda_{\alpha\beta}+\lambda_{\alpha\beta}^{c}\right)\Delta_{\alpha\beta}^{p}\nonumber \\
 & -\sum_{i;\alpha\beta\gamma}\left(\mathcal{D}_{\alpha\beta}R_{\gamma\alpha}+\mathrm{c.c}\right)\left(\Delta^{p}(1-\Delta^{p})\right)_{\gamma\beta}^{1/2}\label{eq:RISB_Lagrangian}.
\end{align}

\noindent Here, $\sum_i$ is shorthand for $\frac{\mathcal{N}_c}{\mathcal{N}}\sum_i$, and in principle, all the variables $\Phi,R,\Delta^{p},E^{c},\lambda,\mathcal{D},\lambda^{c}$ depend on the site index $i$ but we dropped it since we will be looking for uniform solutions.

The slave-boson amplitudes $\Phi_{An}$ appear only in the second
and third lines. Inspired by the fact that these amplitudes are defined
on a local Hilbert space (spanned by $A$) and its copy (spanned by
$n$), one can introduce~\cite{Lanata2015} the corresponding tensor-product
space, spanned by the basis $\{|A\rangle\otimes|n\rangle_a\}_{A,n}$,
where states $|A\rangle$ are created by impurity operators $c^{\dagger},c$
and $|n\rangle_a$ by ``bath'' operators $a^{\dagger},a$. In this
construction, one interprets the amplitudes $\Phi_{An}$ as coefficients
of the Schmidt decomposition of general states $|\Phi\rangle$ of
this product space:
\begin{equation}
|\Phi\rangle\equiv\sum_{An}e^{i\frac{\pi}{2}\mathcal{N}_A(\mathcal{N}_A-1)}\Phi_{An}\hat{U}_{\mathrm{ph}}|A\rangle|n\rangle_a\label{eq:Phi_def},
\end{equation}

\noindent where $\hat{U}_{\mathrm{ph}}$ is a particle-hole transformation acting
only on the $a$ operators. With this definition and the phase factor,
one has~\cite{Lanata2015}:\begin{subequations}
\begin{align}
\mathrm{Tr}\left[F_{\alpha}^{\dagger}F_{\beta}\Phi^{\dagger}\Phi\right] & =\langle\Phi|a_{\beta}a_{\alpha}^{\dagger}|\Phi\rangle\label{eq:Trace_equiv_1},\\
\mathrm{Tr}\left[\Phi^{\dagger}F_{\alpha}^{\dagger}\Phi F_{\beta}\right] & =\langle\Phi|c_{\alpha}^{\dagger}a_{\beta}|\Phi\rangle,\label{eq:Trace_equivalence}
\end{align}
\end{subequations}which allows to express the right-hand sides of Eqs~\eqref{eq:Delta_p_constraint} and \eqref{eq:R_multiplied_constraint} as correlators of the $c$ and $a$ operators. Besides, the second and third lines of the right-hand side of (\ref{eq:RISB_Lagrangian})
become $E^{c}\left(\langle\Phi|\Phi\rangle-1\right)+\langle\Phi|\hat{H}_{\mathrm{embed}}|\Phi\rangle$,
with $H_{\mathrm{embed}}$ defined in Eq. (\ref{eq:H_emb_def}). This
Hamiltonian describes an Anderson impurity level, described by the
fields $c$, $c^{\dagger}$ interacting through the local Hamiltonian
$H_{\mathrm{int}}$, hybridized with noninteracting bath levels $a$,
$a^{\dagger}$ of energies $-\lambda^{c}$ via the hybridization strengths
$\mathcal{D}$. 

Finally, one extremizes the free energy with respect to its variables
to find the Lagrange equations of the problem:\begin{subequations}

\begin{align}
 & \Delta_{\alpha\beta}^{p}=\sum_{\boldsymbol{k}\in\mathrm{BZ},i\omega}\left[i\omega-R\tilde{\varepsilon}_{\boldsymbol{k}}R^{\dagger}-\lambda+\mu\right]_{\beta\alpha}^{-1}e^{i\omega0^{+}},\label{eq:Delta_p-lagrange}\\
 & \sum_{\mu}\left[\left(\Delta^{p}(1-\Delta^{p})\right)^{1/2}\right]_{\alpha\mu}\mathcal{D}_{\beta\mu}\label{eq:D_lagrange}\\
 & \;=\sum_{\boldsymbol{k}\in\mathrm{BZ},i\omega}\left[\left\{ \tilde{\varepsilon}_{\boldsymbol{k}}R^{\dagger}\right\} \left[i\omega-R\tilde{\varepsilon}_{\boldsymbol{k}}R^{\dagger}-\lambda+\mu\right]^{-1}\right]_{\beta\alpha}e^{i\omega0^{+}},\nonumber \\
 & \lambda_{\alpha\beta}^{c}=-\lambda_{\alpha\beta}\nonumber\\
 &\;-\sum_{\gamma\delta\eta}\left\{ \mathcal{D}_{\gamma\delta}R_{\eta\gamma}\frac{\partial\left[\left(\Delta^{p}(1-\Delta^{p})\right)^{1/2}\right]_{\eta\delta}}{\partial\Delta_{\alpha\beta}^{p}}+c.c\right\} ,\label{eq:lambda_c-lagrange}\\
 & \hat{H}_{\mathrm{embed}}|\Phi\rangle=E^{c}|\Phi\rangle,\label{eq:H_imp-1}\\
 & \langle\Phi|a_{\beta}a_{\alpha}^{\dagger}|\Phi\rangle=\Delta_{\alpha\beta}^{p},\label{eq:F1}\\
 & \langle\Phi|c_{\alpha}^{\dagger}a_{\beta}|\Phi\rangle=R_{\gamma\alpha}\left[\left(\Delta^{p}(1-\Delta^{p})\right)^{1/2}\right]_{\gamma\beta}.\label{eq:F2}
\end{align}

\end{subequations}

\subsubsection{Solution of the Lagrange equations\label{subsec:Solution-of-the}}

\paragraph{Root-solving}

In previous works \cite{Lanata2012,Lanata2015,Lanata2016}, the Lagrange
equations were solved by formulating the problem as a root-solving
procedure by defining the functions\begin{subequations}
\begin{align}
\mathcal{F}^{(1)}[R,\lambda] & \equiv\langle\Phi|a_{\beta}a_{\alpha}^{\dagger}|\Phi\rangle-\Delta_{\alpha\beta}^{p}\label{eq:F1_def},\\
\mathcal{F}^{(2)}[R,\lambda] & \equiv\langle\Phi|c_{\alpha}^{\dagger}a_{\beta}|\Phi\rangle-R_{\gamma\alpha}\left[\left(\Delta^{p}(1-\Delta^{p})\right)^{1/2}\right]_{\gamma\beta}.\label{eq:F2_def}
\end{align}
\end{subequations}

\noindent $\mathcal{F}^{(1)}$ and $\mathcal{F}^{(2)}$ are implicit functions
of $R$ and $\lambda$: for a given $R$ and $\lambda$, one can successively
compute $\Delta^{p}$, $\mathcal{D}$, $\lambda^{c}$ and $|\Phi\rangle$
by using Eqs (\ref{eq:Delta_p-lagrange}), (\ref{eq:D_lagrange}),
(\ref{eq:lambda_c-lagrange}) and (\ref{eq:H_imp-1}), respectively.
The fulfillment of the last two equations, Eqs (\ref{eq:F1}) and
(\ref{eq:F2}), thus amounts to solving the root problem:
\begin{align*}
\mathcal{F}^{(1)}[R,\lambda] & =0,\\
\mathcal{F}^{(2)}[R,\lambda] & =0.
\end{align*}

\noindent We note that the same Lagrange equations can be cast as different
root problems (depending on the choice of free variable; for instance,
one could have chosen $\mathcal{D}$ and $\lambda^{c}$ instead of
$R$ and $\lambda$). In the next paragraph, we give an alternative
route to solve the Lagrange equations to shed light on the relation
of RISB with the DMFT method.

\paragraph{Forward recursion and comparison to DMFT}

One can alternatively solve the Lagrange equations in a forward recursive
fashion, as is usually done in DMFT:
\begin{enumerate}
\item Start from a guess $R$ and $\lambda$.
\item Compute the local density matrix $\Delta^{p}$ using (\ref{eq:Delta_p-lagrange})
and the local ``kinetic energy'',
\begin{equation}
\left[\mathcal{K}_{\mathrm{loc}}\right]_{\alpha\beta}\equiv \sum_{\boldsymbol{k}\in\mathrm{BZ},i\omega}\left[\left\{ \tilde{\varepsilon}_{\boldsymbol{k}}R^{\dagger}\right\} \left[i\omega-R\tilde{\varepsilon}_{\boldsymbol{k}}R^{\dagger}-\lambda+\mu\right]^{-1}\right]_{\beta\alpha}e^{i\omega0^{+}}.\label{eq:local_kinetic_energy_def}
\end{equation}
\item Compute $\mathcal{D}$ and $\lambda^{c}$ using Eq. (\ref{eq:D_lagrange}),
i.e
\begin{equation}
\mathcal{D}_{\beta\alpha}=\left[\left(\Delta^{p}(1-\Delta^{p})\right)^{-1/2}\right]_{\alpha\mu}\left[\mathcal{K}_{\mathrm{loc}}\right]_{\mu\beta},\label{eq:D_lagrange-1}
\end{equation}
and Eq. (\ref{eq:lambda_c-lagrange}).
\item Solve the embedded problem Eq (\ref{eq:H_imp-1}) for its (normalized)
ground state $|\Phi\rangle$.
\item Compute a new $R$ as
\begin{equation}
R_{\gamma\alpha}=M_{\alpha\beta}\left[\left(N^{a}(1-N^{a})\right)^{-1/2}\right]_{\beta\gamma}\label{eq:R_from_imp}
\end{equation}
with\begin{subequations}
\begin{align}
N_{\alpha\beta}^{a} & \equiv\langle\Phi|a_{\beta}a_{\alpha}^{\dagger}|\Phi\rangle,\label{eq:N_def}\\
M_{\alpha\beta} & \equiv\langle\Phi|c_{\alpha}^{\dagger}a_{\beta}|\Phi\rangle,\label{eq:M_def}
\end{align}
\end{subequations}and a new $\lambda$ as
\begin{align}
&\lambda_{\alpha\beta}=-\lambda_{\alpha\beta}^{c}\nonumber\\
&\;\;-\sum_{\gamma\delta\eta}\left\{ \mathcal{D}_{\gamma\delta}R_{\eta\gamma}\frac{\partial\left[\left(N^{a}(1-N^{a})\right)^{1/2}\right]_{\eta\delta}}{\partial N_{\alpha\beta}^{a}}+c.c\right\} .\label{eq:lambda_from_imp}
\end{align}
\item Go back to step 2 until convergence of $R$ and $\lambda$.
\end{enumerate}
\begin{figure}
\begin{centering}
\includegraphics[width=0.9\columnwidth]{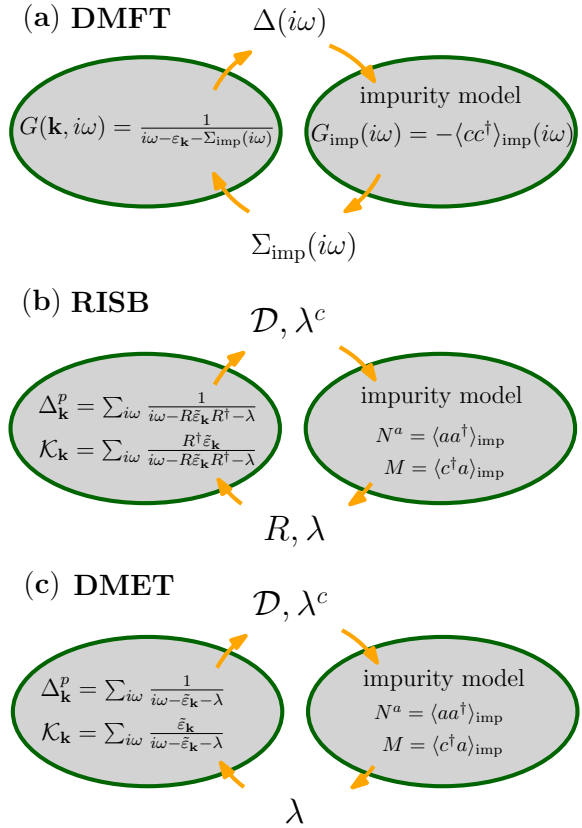}
\par\end{centering}
\caption{From top to bottom: DMFT, RISB and DMET: iterative solution by forward
recursion. (The chemical potential is not indicated for conciseness.)\label{fig:iterative_solution}}
\end{figure}

This cycle is illustrated in Fig. \ref{fig:iterative_solution} (panel (b)): in RISB, the impurity model is solved for $N_{\alpha\beta}$
and $M_{\alpha\beta}$ to obtain the \emph{two} matrices $R_{\alpha\beta}$
and $\lambda_{\alpha\beta}$, which are used as a parametrization
of the lattice self-energy:
\begin{equation}
\Sigma(\boldsymbol{k},i\omega_{n})\approx i\omega_{n}\left(\boldsymbol{1}-(R^{\dagger}R)^{-1}\right)+R^{-1}\lambda\left[R^{\dagger}\right]^{-1}-\boldsymbol{\varepsilon}_{\mathrm{loc}}.\label{eq:Sigma_RISB}
\end{equation}
The impurity model is also parametrized by \emph{two} matrices, the hybridization
strengths $\mathcal{D}_{\alpha\beta}$ and bath hopping parameters
$\lambda_{\alpha\beta}^{c}$, which are adjusted in such a way that
the local density matrix $\Delta^{p}$ coincides with $N$ and $\left[\left(\Delta^{p}(1-\Delta^{p})\right)^{1/2}\right]^{\intercal}R$
coincides with $M$ (Eqs (\ref{eq:F1}-\ref{eq:F2})). 

In practice, this loop allows to obtain stable solutions in the Mott phase of the Hubbard model more easily than by solving the Lagrange equations as a root problem.

By contrast, DMFT (whose self-consistent loop is illustrated in the
top panel of Fig.~\ref{fig:iterative_solution}) requires the \emph{frequency}-\emph{dependent}
local Green's function $G_{\mathrm{loc}}(i\omega_{n})$ (defined as
the $\boldsymbol{k}$-summation of the lattice Green's function $G(\boldsymbol{k},i\omega)$
defined in Eq. (\ref{eq:G_latt_DMFT})) to match the impurity Green's
function $G_{\mathrm{imp}}(i\omega_{n})$ (Eq. (\ref{eq:G_imp_def}))
by adjusting the hybridization \emph{function} $\Delta(i\omega_{n})$,
at the cost of approximating the lattice self-energy by the impurity
self-energy $\Sigma_{\mathrm{imp}}(i\omega_{n})$: all these functions
depend on an infinite number of Matsubara frequencies, and correspondingly
DMFT's impurity model has an infinite number of bath levels: $\Delta(i\omega_{n})$
can in general only be represented by an infinity of bath sites:
\begin{equation}
\Delta_{\alpha\beta}(i\omega_{n})=\sum_{b=1}^{\infty}\sum_{\gamma\delta}\mathcal{D}_{\alpha\gamma}^{b}\left[i\omega\boldsymbol{1}+\left[\lambda^{c}\right]^{b}\right]_{\gamma\delta}^{-1}\left[\mathcal{D}_{\beta\delta}^{b}\right]^{*},\label{eq:hyb_fct_DMFT}
\end{equation}
contrary to RISB in which $b=1$. Note that in the DMFT literature,
$\mathcal{D}$ (resp. $\lambda^{c}$) is usually denoted as $V$ (resp.
$-\epsilon$).

Thus, RISB can be viewed as a well-defined way of drastically truncating
the number of bath levels in the impurity problem, and of parametrizing
the low-energy behavior of the impurity self-energy by two observables,
the matrices $R$ and $\lambda$. Beyond the reduced number of bath
levels of the impurity model, RISB only necessitates the computation
of \emph{static} correlators, $\langle\Phi|a_{\beta}a_{\alpha}^{\dagger}|\Phi\rangle$
and $\langle\Phi|c_{\alpha}^{\dagger}a_{\beta}|\Phi\rangle$, whereas
DMFT requires the full frequency dependence of the Green's function
$G_{\mathrm{imp}}(i\omega_{n})$.

Alternative approaches to truncate the number of bath levels exist:
two-site DMFT \cite{Potthoff2001} uses
the low and high-frequency limit of the DMFT self-consistency condition
$G_{\mathrm{imp}}(i\omega)=G_{\mathrm{loc}}(i\omega)$ to fix the
position and hybridization of a single bath level (in the context
of a single-band model; see also \cite{Kolorenc2012,Kolorenc2015}
for another prescription in a multiorbital context).

In a different perspective, solving the DMFT impurity model with an
exact diagonalization method \cite{Caffarel1994} also relies on a truncation procedure.
There, the number as well as position and hybridization of bath sites
is dictated not by formal considerations (as in RISB or two-site DMFT)
but by computational limitations attached to the former and a (somewhat
arbitrary) fitting procedure (using the parametrization of Eq. (\ref{eq:hyb_fct_DMFT}))
for the latter.

\subsubsection{Equivalence to the multiband Gutzwiller approximation}

RISB has been shown to be equivalent~\cite{Bunemann2007} to the multiband
formulation of the Gutzwiller approximation. In other words, the above
derivation can be carried out, instead of introducing slave bosons
$\phi$ and quasiparticle fields $f$, by minimizing the variational
energy $\langle\Psi_{\mathrm{G}}|H|\Psi_{\mathrm{G}}\rangle$ over
the Gutzwiller wavefunctions
\begin{equation}
|\Psi_{\mathrm{G}}(R,\lambda)\rangle\equiv\prod_{i}\mathcal{P}_{i}(R,\lambda)|\Psi_{0}(R,\lambda)\rangle,\label{eq:Psi_Gutz}
\end{equation}

\noindent where $|\Psi_{0}\rangle$ is a noninteracting wavefunction (a Slater
determinant $|\Psi_{0}\rangle=\prod_{p=1}^{n_{\mathrm{occ}}}c_{p}^{\dagger}|0\rangle$
with $p=1\dots n_{\mathrm{occ}}$ denoting the occupied states) and
$\mathcal{P}_{i}$ is a projector on the local many-body Hilbert space
defined above:
\begin{equation}
\mathcal{P}_i = \sum_{A,B} \Lambda_{A,B}(R,\lambda) |A_i\rangle \langle B_i|
\end{equation}
(the connection between the $\Lambda$ matrix and the slave-boson matrix $\Phi_{An}$ is explored \emph{e.g} in Ref.~\onlinecite{Lanata2012}). 

This is done under the ``Gutzwiller constraints''\begin{subequations}
\begin{align}
\forall i\;\;\; & \langle\Psi_{0}|\mathcal{P}_{i}^{\dagger}\mathcal{P}_{i}|\Psi_{0}\rangle=1,\label{eq:Gutz_constraint_1}\\
\forall i\;\;\; & \langle\Psi_{0}|\mathcal{P}_{i}^{\dagger}\mathcal{P}_{i}c_{i\alpha}^{\dagger}c_{i\beta}|\Psi_{0}\rangle=\langle\Psi_{0}|c_{i\alpha}^{\dagger}c_{i\beta}|\Psi_{0}\rangle.\label{eq:Gutz_constraint_2}
\end{align}
\end{subequations}which can be shown (see e.g. Ref.~\onlinecite{Bunemann2007}) to be equivalent to the aforementioned
RISB constraints, Eqs~\eqref{eq:constraint_1-1}-\eqref{eq:constraint_2-1}.

In the next section, we show that the recently introduced DMET is
a simplified form of RISB where $R$ is approximated by the identity
matrix.

\section{Density-matrix embedding theory: a simplified RISB\label{sec:Density-matrix-embedding-theory:}}

Density-matrix embedding theory (DMET) has been introduced as a simplified
version of DMFT~\cite{Knizia2012}. As outlined in section \ref{sec:Overview},
DMET replaces the original lattice problem (Eq. (\ref{eq:Hamiltonian-with-H-loc}))
with a reference problem (Eq. (\ref{eq:H_DMET})) where the effect
of correlations is described by a one-body potential $\lambda$ (usually
called $u$ in the DMET literature). In other words, the self-energy
is approximated by a static, local (within a cell) potential 
\begin{equation}
\Sigma_{\alpha\beta}(\boldsymbol{k},i\omega_{n})\approx u_{\alpha\beta}\label{eq:Sigma_DMET}
\end{equation}

Like in DMFT and RISB, the approximate form of the self-energy (here
a matrix, $u$) is obtained in a nontrivial way by a self-consistent
mapping of the reference problem onto an embedded local problem with
fewer degrees of freedom. The parameters of this impurity problem
are then adjusted (through an adjustment of $u$) to match the (one-particle)
density matrix of the embedded problem with the projection of the
density matrix of the reference problem onto the embedded subspace. 

In a wavefunction language, DMET corresponds to making the following
variational ansatz for the ground-state wavefunction:

\begin{equation}
|\Psi_{\mathrm{DMET}}\rangle\equiv|\Psi_{0}(u)\rangle,\label{eq:wavefct_ansatz_DMET}
\end{equation}

\noindent where $|\Psi_{0}\rangle$ is a Slater determinant and $u$ is obtained
by matching the density-matrices as discussed above.

Several equivalent derivations of DMET have been given in the literature~\cite{Knizia2012,Knizia2013,Zheng2016,Zheng2017,Bulik2014,Bulik2015,Wouters2016,Wouters2016a}.
Here, we give a derivation similar to \cite{Bulik2014} but at times
use different notation or terminology to make the connection to the
RISB or Gutzwiller formalism more transparent.

\subsection{Summary of the DMET self-consistency}

In this subsection, we give a brief summary of the DMET workflow.
Having mapped the lattice problem onto a reference problem consisting
in an interacting \emph{fragment} (of size $\mathcal{N}_c$) and a noninteracting
\emph{environment} (as illustrated in Fig. \ref{fig:DMET-fragment-environment}),
DMET defines an embedded subspace through a projection operator $P$
which projects the $\mathcal{N}_c+(\mathcal{N}-\mathcal{N}_c)$ degrees of freedom
of the fragment + environment onto the $\mathcal{N}_c+\mathcal{N}_c$ degrees of freedom
of the embedded (i.e impurity+bath) problem. Then, the DMET self-consistency
consists in matching:

(i) \emph{the density matrix of the embedded problem} 
\begin{equation}
\rho_{\mathrm{imp}}\equiv\left[\begin{array}{cc}
\langle\Phi|c_{\alpha}^{\dagger}c_{\beta}|\Phi\rangle & \langle\Phi|c_{\alpha}^{\dagger}a_{\beta}|\Phi\rangle\\
\langle\Phi|a_{\alpha}^{\dagger}c_{\beta}|\Phi\rangle & \langle\Phi|a_{\alpha}^{\dagger}a_{\beta}|\Phi\rangle
\end{array}\right],\label{eq:rho_imp_def}
\end{equation}
\noindent where $|\Phi\rangle$ is the ground-state of $H_{\mathrm{embed}}$
(Eq. (\ref{eq:H_emb_def})), itself related to $H_{\mathrm{eff}}$ (Eq.~(\ref{eq:H_DMET}))
by ``projection'' by $P$, namely
\begin{equation}
H_{\mathrm{embed}}=\hat{H}_{\mathrm{loc}}[\{c_{\alpha}^{\dagger},c_{\alpha}\}]+\sum_{\alpha\beta}\left[\begin{array}{c}
c_{\alpha}\\
a_{\alpha}
\end{array}\right]^{\dagger}\left[h_{\mathrm{embed}}\right]_{\alpha\beta}\left[\begin{array}{c}
c_{\beta}\\
a_{\beta}
\end{array}\right],\label{eq:H_embed_as_projection}
\end{equation}
\noindent with
\begin{align}
h_{\mathrm{embed}} & \equiv P^{\dagger}hP,\label{eq:h_embed_projection}
\end{align}
and $h$ the one-body part of $H_{\mathrm{eff}}$:
\begin{equation}
h\equiv t + \left[\begin{array}{cc}
0 & 0\\
0 & \left[\begin{array}{ccc}
u\\
 & \ddots\\
 &  & u
\end{array}\right] 
\end{array}\right],\label{eq:h_def}
\end{equation}

(the $u$ block is of size $(\mathcal{N}-\mathcal{N}_c)\times (\mathcal{N}-\mathcal{N}_c)$)

\noindent with

(ii) \emph{the projection of the density matrix $\rho$ of the reference
problem} in the embedded subspace:
\begin{equation}
\rho_{\mathrm{embed}}\equiv P^{\dagger}\rho P,\label{eq:rho_embed_def}
\end{equation}
with
\begin{equation}
\rho_{i\alpha,j\beta}=\sum_{i\omega}e^{i\omega0^{+}}\left[i\omega\boldsymbol{1}-h_\mathrm{TI}+\mu\right]_{i\alpha,j\beta}^{-1},\label{eq:rho_noninteracting}
\end{equation}
We recall that the Latin indices $i,j$ label the unit cells, while the Greek indices $\alpha,\beta=1\dots \mathcal{N}_c$ label the internal orbital/cluster degrees of freedom. 
In the above expression, $h_\mathrm{TI}$ is the translation-invariant version of $h$, i.e it contains a potential $u$ on the upper-left (i=0,j=0) block: $[h_\mathrm{TI}]_{i\alpha,j\beta}=h_{i\alpha,j\beta}+u_{\alpha\beta}\delta_{i,0}\delta_{j,0}$.
In practice, we will see that exactly \emph{matching} those two density
matrices is in general impossible, so that a \emph{minimization} of
a distance between both matrices is carried out.

In the following subsection, we show how the projector $P$ is constructed.

\subsection{Construction of the impurity and bath operators}

In this section, we define the mapping from the effective medium (with
$\mathcal{N}$ orbitals) to the embedded problem (with $2\mathcal{N}_c$ orbitals), or in other words we construct the projector $P$. 

We start by explaining how to transform from the fragment (size $\mathcal{N}_c$) to the impurity (size $\mathcal{N}_c$) orbitals.
First off, diagonalizing the lattice density matrix $\rho$ (given in
Eq. (\ref{eq:rho_noninteracting})) yields a transformation $D_{\mathrm{F}}^{\mathrm{occ}}$
from the single-site levels of the fragment (denoted by Greek indices)
to the occupied levels of $H_\mathrm{eff}$ (see Appendix \ref{sec:Transformation-of-the}
for details). 

Second, we need to find a transformation from the occupied levels to the impurity orbitals. The central object
for doing so is the overlap matrix between the fragment and
the occupied states of the lattice, defined as: 
\begin{equation}
S_{pq}^{\mathrm{occ}}\equiv\langle\phi_{p}|P^{\mathrm{F}}|\phi_{q}\rangle,\label{eq:S_occ_def}
\end{equation}
\noindent with $p$ and $q$ labelling two occupied states of $H_\mathrm{eff}$ ($1\leq p,q\leq n_{\mathrm{occ}}$), and $\phi_p$ the corresponding single-particle state: $|\phi_{p}\rangle\equiv c_{p}^{\dagger}|0\rangle$, and $P^{\mathrm{F}}$
is the projector on the fragment ($P^{\mathrm{F}}\equiv\sum_{\alpha=1}^{\mathcal{N}_c}|\phi_\alpha\rangle\langle \phi_\alpha|$).
We show in Appendix \ref{sec:Transformation-of-the} that this matrix
can be transformed to a diagonal form:
\begin{equation}
S^{\mathrm{occ}}=V_{\mathrm{F}}n^{0}V_{\mathrm{F}}^{\dagger},\label{eq:S_occ_expr}
\end{equation}

\noindent where $n^{0}$ is a $\mathcal{N}_c\times \mathcal{N}_c$ diagonal matrix, and $V_{\mathrm{F}}$
is a $n_{\mathrm{occ}}\times \mathcal{N}_c$ rectangular matrix such that
$V_{\mathrm{F}}^{\dagger}V_{\mathrm{F}}=1$. $\left[V_{\mathrm{F}}\right]_{p,\alpha'}$
defines a transformation from the occupied states ($p,\dots$) to
new states (denoted by primed Greek indices $\alpha',\dots$) that
correspond to the ``natural orbitals'' used e.g. in \cite{Lanata2012}. 

With these two transformations, one defines the transformation
from the fragment to the impurity as:
\[
\tilde{C}_{\mathrm{F}}=D_{\mathrm{F}}^{\mathrm{occ}}V_{\mathrm{F}},
\]

\noindent or rather, with orthonormalized columns:

\begin{equation}
\left[C_{\mathrm{F}}\right]_{\alpha\alpha'}\equiv\frac{\left[D_{\mathrm{F}}^{\mathrm{occ}}\right]_{\alpha p}\left[V_{\mathrm{F}}\right]_{p\alpha'}}{\sqrt{n_{\alpha'}^{0}}},\label{eq:C_F_def}
\end{equation}

\noindent (to determine the normalization, we have used: $\tilde{C}_{\mathrm{F}}^{\dagger}\tilde{C}_{\mathrm{F}}=V_{\mathrm{F}}^{\dagger}D_{\mathrm{F}}^{\mathrm{occ},\dagger}D_{\mathrm{F}}^{\mathrm{occ}}V_{\mathrm{F}}=V_{\mathrm{F}}^{\dagger}V_{\mathrm{F}}n^{0}V_{\mathrm{F}}^{\dagger}V_{\mathrm{F}}=n^{0}$). 

Likewise, the matrix which projects from the environment to the bath
is defined as the product of the transformation from the environment
levels to the occupied states (a matrix called $D_{\mathrm{E}}^{\mathrm{occ}}$)
with the transformation from the occupied levels to the natural orbitals
($V_{\mathrm{F}}$). After orthonormalization of the columns, we obtain:

\begin{equation}
C_{\mathrm{B}}\equiv\frac{D_{\mathrm{E}}^{\mathrm{occ}}V_{\mathrm{F}}}{\sqrt{1-n^{0}}}.\label{eq:C_B_def}
\end{equation}

\noindent We thus define the projector:
\begin{equation}
\overline{P}\equiv\left[\begin{array}{cc}
C_{\mathrm{F}}\\
 & C_{\mathrm{B}}
\end{array}\right],\label{eq:P_def}
\end{equation}

\noindent which projects the lattice problem (fragment $\lbrace c_{1\alpha}^{\dagger}\rbrace_{1\leq\alpha\leq \mathcal{N}_c}$+environment
$\lbrace c_{i\alpha}^{\dagger}\rbrace_{1<i\leq\mathcal{N}/\mathcal{N}_c,1\leq\alpha\leq \mathcal{N}_c}$)
onto the embedded problem (impurity $\lbrace\tilde{c}_{\alpha'}^{\dagger}\rbrace_{1\leq\alpha'\leq \mathcal{N}_c}$+bath
$\lbrace\tilde{a}_{\alpha'}^{\dagger}\rbrace_{1\leq\alpha'\leq \mathcal{N}_c}$):
\begin{align}
\tilde{c}_{\alpha'}^{\dagger} & =\sum_{\alpha=1}^{\mathcal{N}_c}c_{1\alpha}^{\dagger}\left[C_{\mathrm{F}}\right]_{\alpha\alpha'},\label{eq:imp_creation_op}\\
\tilde{a}_{\alpha'}^{\dagger} & =\sum_{i=2}^{\mathcal{N}/\mathcal{N}_c}\sum_{\alpha=1}^{\mathcal{N}_c}c_{i\alpha}^{\dagger}\left[C_{\mathrm{B}}\right]_{i\alpha;\alpha'}.\label{eq:bath_creation_op}
\end{align}

Instead of the natural-orbital basis, one can choose instead to use
the original basis (denoted by unprimed Greek indices) as a single-site
basis to express the creation operators $c^{\dagger}$ and $a^{\dagger}$
of the embedded problem. This is done by defining the alternative
projector:
\begin{equation}
P\equiv\left[\begin{array}{cc}
\boldsymbol{1}\\
 & C_{\mathrm{B}}C_{\mathrm{F}}^{\dagger}
\end{array}\right],\label{eq:Pbar_def}
\end{equation}

\noindent which is related to $\overline{P}$ by a unitary transform,
\[
P=\overline{P}\left[\begin{array}{cc}
C_{\mathrm{F}}^{\dagger}\\
 & C_{\mathrm{F}}^{\dagger}
\end{array}\right],
\]

\noindent where $\overline{P}$ projects into the natural orbitals, while $P$ projects
into the original orbitals.

We note that the above construction corresponds to carrying out the Schmidt decomposition of $|\Psi_0(u)\rangle$~\cite{Knizia2012}.

In the next subsection, we use $P$ to
project lattice observables onto the embedded subspace.

\subsection{Projections in the embedded subspace\label{subsec:Projections-in-the}}

After constructing the impurity and bath levels, one can now map the
density matrix and lattice Hamiltonian onto the embedded subspace.

\subsubsection{Embedded density matrix}

The projection of the density matrix onto the embedded subspace is
defined in Eq. (\ref{eq:rho_embed_def}), and similarly for $\overline{\rho}_{\mathrm{embed}}$.
After a few algebraic steps detailed in Appendix \ref{subsec:Density-matrix},
we obtain 
\begin{equation}
\overline{\rho}_{\mathrm{embed}}\equiv\overline{P}^{\dagger}\rho\overline{P}=\left[\begin{array}{cc}
n^{0} & \sqrt{n^{0}(1-n^{0})}\\
\sqrt{n^{0}(1-n^{0})} & 1-n^{0}
\end{array}\right],\label{eq:rho_embed_no}
\end{equation}

\noindent i.e. $\overline{\rho}_{\mathrm{embed}}$ is entirely determined by
the occupations of the natural orbitals $n^{0}$. Similarly, its expression
in the original basis is
\begin{equation}
\rho_{\mathrm{embed}}=\left[\begin{array}{cc}
\Delta^{p} & \sqrt{\Delta^{p}(1-\Delta^{p})}\\
\sqrt{\Delta^{p}(1-\Delta^{p})} & 1-\Delta^{p}
\end{array}\right],\label{eq:rho_embed}
\end{equation}

\noindent with
\[
\Delta^{p}\equiv C_{\mathrm{F}}n^{0}C_{\mathrm{F}}^{\dagger}=D_{F}^{\mathrm{occ}}D_{\mathrm{F}}^{\mathrm{occ},\dagger}=\rho_{\mathrm{F}},
\]

\noindent with $\rho_{\mathrm{F}}$ the top-left $\mathcal{N}_c\times \mathcal{N}_c$ block
of $\rho$. Thus, $\Delta^{p}$ (as defined in the RISB section) is
the one-particle density matrix of the fragment, $\rho_{\mathrm{F}}$.
Using (\ref{eq:rho_noninteracting}), we thus have:
\begin{align*}
\Delta_{\alpha\beta}^{p} & =\sum_{i\omega}e^{i\omega0^{+}}\left[i\omega\boldsymbol{1}-t-u+\mu\right]_{1\alpha,1\beta}^{-1}\\
 & =\sum_{\boldsymbol{k},i\omega}e^{i\omega0^{+}}\left[i\omega\boldsymbol{1}-\varepsilon_{\boldsymbol{k}}-u+\mu\right]_{\alpha,\beta}^{-1}.
\end{align*}

We define
\begin{equation}
\tilde{u}_{\alpha\beta}\equiv u_{\alpha\beta}+\left[\varepsilon_{\mathrm{loc}}\right]_{\alpha\beta}\label{eq:u_tilde_def}
\end{equation}

\noindent to obtain the analog of Eq. (\ref{eq:Delta_p-lagrange}) in the RISB
formalism:
\begin{equation}
\Delta_{\alpha\beta}^{p}=\sum_{\boldsymbol{k},i\omega}e^{i\omega0^{+}}\left[i\omega\boldsymbol{1}-\tilde{\varepsilon}_{\boldsymbol{k}}-\tilde{u}+\mu\right]_{\alpha,\beta}^{-1}.\label{eq:Delta_p_DMET}
\end{equation}

\noindent Eqs (\ref{eq:Delta_p-lagrange}) and (\ref{eq:Delta_p_DMET}) can
be identified provided:\begin{subequations}
\begin{align}
R & =\boldsymbol{1},\label{eq:R_DMET}\\
\lambda & =\tilde{u}.\label{eq:lambda_DMET}
\end{align}
\end{subequations}In the next subsection, we show that this identification
holds for all other DMET observables.

\subsubsection{Parameters of the embedded problem}

Based on the two definitions of $H_{\mathrm{embed}}$, Eqs. (\ref{eq:H_emb_def})
and (\ref{eq:H_embed_as_projection}), we can write\footnote{The minus sign in front of $\lambda^{c}$ stems from the fact that
$H_{\mathrm{embed}}$ contains a term $\lambda^{c}aa^{\dagger}$ instead
of the more familiar $\lambda^{c}a^{\dagger}a$}:
\begin{equation}
h_{\mathrm{embed}}=\left[\begin{array}{cc}
t_{\mathrm{F}} & \mathcal{D}\\
\mathcal{D}^{\dagger} & -\lambda^{c}
\end{array}\right].\label{eq:h_embed_parameters}
\end{equation}

\noindent Identifying the right-hand sides of Eqs (\ref{eq:h_embed_projection})
and (\ref{eq:h_embed_parameters}), and thanks to the definition {[}Eq.
(\ref{eq:Pbar_def}){]} of $P$, one can show, after a few algebraic
steps detailed in Appendix \ref{subsec:Parameters-of-the}, that:\begin{subequations}

\begin{equation}
\mathcal{D}=\sum_{\boldsymbol{k}i\omega}\boldsymbol{\tilde{\varepsilon}_{\boldsymbol{k}}}\left[i\omega-\tilde{\boldsymbol{\varepsilon}}_{\boldsymbol{k}}-\tilde{u}\right]^{-1}\left[\sqrt{\Delta^{p}(1-\Delta^{p})}\right]^{-1},\label{eq:D_final_DMET}
\end{equation}

\noindent and

\begin{align}
 & \lambda^{c}=-\tilde{u}-\left[\sqrt{(1-\Delta^{p})\Delta^{p}}\right]^{-1}\label{eq:lambda_c_final}\\
 & \;\;\;\;\;\;\times\frac{\mathcal{K}_{\mathrm{loc}}(1-2\Delta^{p})+(1-2\Delta^{p})\mathcal{K}_{\mathrm{loc}}}{2}\left[\sqrt{\Delta^{p}(1-\Delta^{p})}\right]^{-1}\nonumber 
,\end{align}
\end{subequations}which respectively correspond to Eqs. (\ref{eq:D_lagrange})
and (\ref{eq:lambda_c-lagrange}) with the identification (\ref{eq:R_DMET}-\ref{eq:lambda_DMET}). 

\subsection{Self-consistency conditions}

As mentioned in a previous section, the DMET self-consistency conditions
consist in matching the embedded density matrix $\rho_{\mathrm{embed}}$
obtained by projection of the lattice-density matrix onto the embedded
subspace with the density matrix of the embedded or impurity problem,
whose block structure reads:
\begin{equation}
\rho_{\mathrm{imp}}=\left[\begin{array}{cc}
N^{c} & M\\
M^{\dagger} & N^{a}
\end{array}\right].\label{eq:rho_imp_blocks}
\end{equation}

\noindent with $N^{a}$ and $M$ defined in Eqs (\ref{eq:N_def}-\ref{eq:M_def})
and \begin{align}
N_{\alpha\beta}^{c} & \equiv\langle\Phi|c_{\alpha}^{\dagger}c_{\beta}|\Phi\rangle,\label{eq:Nc_def}
\end{align}
\noindent where $|\Phi\rangle$ is the ground state of the
impurity Hamiltonian, i.e the solution of Eq.(\ref{eq:H_imp-1}).
Thus, the self-consistency conditions explicitly read\begin{subequations}
\begin{align}
\langle\Phi|a_{\beta}a_{\alpha}^{\dagger}|\Phi\rangle & =\Delta_{\alpha\beta}^{p}\label{eq:DMET_sc_1},\\
\langle\Phi|c_{\alpha}^{\dagger}a_{\beta}|\Phi\rangle & =\left[\sqrt{\Delta^{p}(1-\Delta^{p})}\right]_{\alpha\beta},\label{eq:DMET_sc_2}\\
\langle\Phi|c_{\alpha}^{\dagger}c_{\beta}|\Phi\rangle & =\Delta_{\alpha\beta}^{p}.\label{eq:DMET_sc_3}
\end{align}
\end{subequations}The first two lines, with the identification (\ref{eq:R_DMET}),
correspond to the RISB conditions (\ref{eq:F1}-\ref{eq:F2}).

\subsection{Solution of the DMET equations: overdetermination, idempotency and
alternative self-consistency conditions}

The DMET equations presented in the previous sections have so far
been solved in a forward recursive way:
\begin{enumerate}
\item Start from a guess for $u$.
\item Compute $\mathcal{D}$ and $\lambda^{c}$ from $\Delta^{p}$ and $\mathcal{K}_{\mathrm{loc}}$.
\item Solve the impurity model for $\rho_{\mathrm{imp}}$, i.e for $N^{a}$,
$N^{c}$ and $M$.
\item From $u$, compute $\rho_{\mathrm{embed}}(u)$ as given by Eq. (\ref{eq:rho_embed}).
If $\rho_{\mathrm{embed}}(u)=\rho_{\mathrm{imp}}(u)$, self-consistency
is reached and the solution is $u$. Otherwise, find $u'$ such that
$\rho_{\mathrm{embed}}(u')=\rho_{\mathrm{imp}}(u)$ and go back to
step 2 with the new $u'$ until self-consistency is reached.
\end{enumerate}
This loop is different from the
loop presented in section \ref{subsec:Solution-of-the}. The potential advantage of this alternative forward recursion is that it
in principle requires fewer computations of the impurity solution:
the root problem,
\begin{equation}
\mathcal{F}_{u}(u')\equiv\rho_{\mathrm{embed}}(u')-\rho_{\mathrm{imp}}(u)=0,\label{eq:root_DMET}
\end{equation}

\noindent requires only one impurity computation (to compute $\rho_{\mathrm{imp}}(u)$). However, this root problem must be solved several times, so that the numerical gain is \emph{a priori} unclear.

On the other hand, the DMET self-consistency condition leads to an
overdetermined root problem: there is only one unknown $u$ to satisfy
three self-consistency conditions Eqs (\ref{eq:DMET_sc_1}-\ref{eq:DMET_sc_2}-\ref{eq:DMET_sc_3}).
In comparison, the root problem to be solved in RISB has as many unknowns
($R$ and $\lambda$) as equations ($\mathcal{F}^{(1)}=0$ and $\mathcal{F}^{(2)}=0$).
Another independent issue is that $\rho_{\mathrm{embed}}$ as given
in Eq. (\ref{eq:rho_embed_no}) or (\ref{eq:rho_embed}) is idempotent
(one can check that $\rho_{\mathrm{embed}}^{2}=\rho_{\mathrm{embed}}$),
with the consequence that its eigenvalues must be zero or one. That
$\rho_{\mathrm{imp}}$ generically shares this property is improbable;
in fact, converged RISB results in the literature (with $R\neq1$)
prove that $\rho_{\mathrm{imp}}$ is in general not idempotent.

This has led to the exploration of several (arbitrary) procedures
in the literature: the original papers proposed to minimize the sum
of the squared differences between the matrix elements of $\rho_{\mathrm{embed}}$
and $\rho_{\mathrm{imp}}$ (instead of trying to find the root of
Eq. (\ref{eq:root_DMET})); other authors suggest to fulfill only
the condition on the density (e.g. Eq (\ref{eq:DMET_sc_1})), a scheme
dubbed ``density embedding theory'', DET \cite{Bulik2014}.

In Fig. \ref{fig:iterative_solution} (bottom panel), we illustrate
another possible recursive scheme to solve the DMET equations inspired
from DMFT (this scheme corresponds to the forward recursion presented
in section \ref{subsec:Solution-of-the}, only with $R=1$). This
figure, while emphasizing the similarities between the three methods,
also hints at the overdetermination problem we just discussed: while
in RISB, two observables are needed to compute ($N^{a}$ and $M$)
and parametrize ($\lambda$ and $R$) the self-energy \emph{and} to
characterize the embedded problem ($\lambda^{c}$ and $\mathcal{D}$),
in DMET, two observables ($N^{a}$ and $M$) are
computed at the level of the embedded problem (and needed to describe
it, $\lambda^{c}$ and $\mathcal{D}$), but the self-energy is described
by only one parameter ($\lambda$ or $u$), possibly pointing to an underexploitation
of the physical information contained in the solution of the impurity model.

\section{Conclusion\label{sec:Conclusion}}

In this work, we have derived the relation between two methods, RISB
and DMET, which can both be regarded as simplified versions of DMFT.
As such, they can access regimes of parameters and systems for which
the exact solution, via quantum Monte-Carlo, of the DMFT impurity
problem, is prohibitively costly if not out of reach due to the negative
sign problem or very large computing times.

We have shown that the DMET equations can be obtained from the RISB
equations by setting the quasiparticle weight factor to 1 in RISB.
This allows to establish a clear connection between these two methods, which
are both based on the mapping of a strongly correlated problem onto
a simplified problem describing correlated orbitals embedded in a
noninteracting host.  

An  additional  comparison among the methods is possible if  one uses the interpretation of the RISB method as a linear expansion of a self-energy \cite{Lanata2015}. Therefore, if one focusses on the the low-energy behavior of the self-energy, DMFT has real and imaginary parts with a general frequency dependence, RISB keeps the constant and linear term in a real self-energy, and DMET is purely static. In this context, it is worth mentioning  other approximate methods which use a very different parametrization
of the self energy in terms of a continuous fraction expansion (see for instance Refs.~\onlinecite{Onoda2003,Avella2006}).

This common perspective on the three methods naturally suggests transposing extensions of one method to the others. For instance, a simple generalization of the RISB/Gutzwiller ground-state energy to a temperature-dependent free energy, briefly exposed in Appendix \ref{sec:gs_energy}, can be used to derive a temperature-dependent DMET free energy.

This work opens  additional questions for cluster extensions. 
For instance, DMET yields
 good spectra for the Hubbard model~\cite{Booth2015}.   Given that RISB and DMET
have the same computational cost (that of solving an impurity model
with the same number of bath and impurity levels),
similar calculations should be carried out with the RISB method to explore how that
embedding accelerates the convergence to  the thermodynamic limit for spectral properties.

Applications of DMFT to molecular systems already exist~\cite{Jacob2010,Lin2011}, but it has been difficult to extend it to complex molecules. 
On the other hand,  DMET  has been very successful in its applications to quantum chemistry~\cite{Wouters2016a}. It would be interesting
to explore potential applications of RISB in that field as well.

\begin{acknowledgments}
This project is supported by the Department of Energy under grant
DE-FG02-99ER45761.
\end{acknowledgments}

\appendix

\section{Transformation of the overlap matrix to diagonal form\label{sec:Transformation-of-the}}

We start by diagonalizing the noninteracting Hamiltonian $h$; we
obtain:
\[
h=D\epsilon D^{\dagger},
\]

\noindent with $D$ a $\mathcal{N}\times\mathcal{N}$ unitary matrix and $\epsilon=\mathrm{diag}(\{\epsilon_{k}\}_{k=1\dots})$
a diagonal matrix. Thanks to the expression (\ref{eq:rho_noninteracting}),
$D$ also diagonalizes the density matrix, \emph{i.e.}
\begin{align}
\rho & =DnD^{\dagger}\label{eq:rho_diag},
\end{align}

\noindent with $n$ a diagonal matrix with entries $=n_{\mathrm{F}}(\epsilon_{k})$.
The first $n_{\mathrm{occ}}$ eigenvalues of $\rho$ (i.e the first
$n_{\mathrm{occ}}$ entries of $n$) are unity (they correspond to
the occupied states), while the other eigenvalues vanish (they correspond
to the empty states).

We now split $D$ into its fragment and its environment blocks:

\begin{equation}
D=\left[\begin{array}{c}
D_{\mathrm{F}}\\
D_{\mathrm{E}}
\end{array}\right],\label{eq:D_F_E}
\end{equation}

\noindent where $D_{\mathrm{F}}$ is of size $\mathcal{N}_c\times\mathcal{N}$. Since
$D$ is unitary, the following properties hold:
\begin{align}
D_{\mathrm{F}}^{\dagger}D_{\mathrm{F}}+D_{\mathrm{E}}^{\dagger}D_{\mathrm{E}} & =1,\label{eq:D_prop_1}\\
D_{\mathrm{F}}D_{\mathrm{F}}^{\dagger} & =1,\nonumber \\
D_{\mathrm{E}}D_{\mathrm{E}}^{\dagger} & =1\nonumber. 
\end{align}
\noindent We further decompose $D_{\mathrm{F}}$ into two blocks
\begin{equation}
D_{\mathrm{F}}=\left[\begin{array}{cc}
D_{\mathrm{F}}^{\mathrm{occ}} & D_{\mathrm{F}}^{\mathrm{unocc}}\end{array}\right],\label{eq:D_F_occ}
\end{equation}

\noindent with $D_{\mathrm{F}}^{\mathrm{occ}}$ a $\mathcal{N}_c\times n_{\mathrm{occ}}$
matrix. Note that (\ref{eq:D_prop_1}) implies:
\begin{equation}
D_{\mathrm{F}}^{\mathrm{occ}\dagger}D_{\mathrm{F}}^{\mathrm{occ}}+D_{\mathrm{E}}^{\mathrm{occ}\dagger}D_{\mathrm{E}}^{\mathrm{occ}}=1.\label{eq:D_occ_relations}
\end{equation}

\noindent We now perform a singular value decomposition of $D_{\mathrm{F}}^{\mathrm{occ}}$.
We obtain:
\begin{equation}
D_{\mathrm{F}}^{\mathrm{occ}}=U\left\{ \mathrm{diag}\{\sqrt{n^{0}}\},0\right\} V^{\dagger},\label{eq:SVD_D_F_occ}
\end{equation}

\noindent with $U$ a $\mathcal{N}_c\times \mathcal{N}_c$ unitary matrix, $\left\{ \mathrm{diag}\{\sqrt{n^{0}}\},0\right\} $
a $\mathcal{N}_c\times n_{\mathrm{occ}}$ matrix (with $\mathrm{diag}\{\sqrt{n^{0}}\}$
a $\mathcal{N}_c\times \mathcal{N}_c$ diagonal matrix, simply denoted as $\sqrt{n^{0}}$
in the following), and $V$ a $n_{\mathrm{occ}}\times n_{\mathrm{occ}}$
unitary matrix. We decompose $V$ into two blocks:
\begin{equation}
V=\left[\begin{array}{cc}
V_{\mathrm{F}} & V_{\mathrm{E}}\end{array}\right],\label{eq:V_decomp}
\end{equation}

\noindent with $V_{\mathrm{F}}$ a $n_{\mathrm{occ}}\times \mathcal{N}_c$ matrix. The
unitarity of $V$ implies the properties
\begin{align}
V_{\mathrm{F}}V_{\mathrm{F}}^{\dagger}+V_{\mathrm{E}}V_{\mathrm{E}}^{\dagger} & =1\nonumber, \\
V_{\mathrm{F}}^{\dagger}V_{\mathrm{F}} & =1,\label{eq:VF_rel}\\
V_{\mathrm{E}}^{\dagger}V_{\mathrm{E}} & =1.\nonumber 
\end{align}
Plugging (\ref{eq:V_decomp}) into (\ref{eq:SVD_D_F_occ}), we obtain:

\begin{equation}
D_{\mathrm{F}}^{\mathrm{occ}}=U\sqrt{n^{0}}V_{\mathrm{F}}^{\dagger}\label{eq:Docc_rel}
\end{equation}

\noindent The last step is to notice that the overlap matrix $S^{\mathrm{occ}}$,
defined in Eq. (\ref{eq:S_occ_def}), is also given by the expression
\begin{equation}
S^{\mathrm{occ}}=D_{\mathrm{F}}^{\mathrm{occ},\dagger}D_{\mathrm{F}}^{\mathrm{occ}}.\label{eq:S_occ_expr-1}
\end{equation}

\noindent Thus, using (\ref{eq:Docc_rel}), we obtain Eq. (\ref{eq:S_occ_expr}).

\section{Projections into the embedded subspace}

We start by noting that the transformation between site indices and
natural orbital indices is given by $C_{\mathrm{F}}$ (defined in
Eq (\ref{eq:C_F_def})), itself equal to $U$:
\begin{equation}
C_{\mathrm{F}}=\frac{U\sqrt{n^{0}}V_{\mathrm{F}}^{\dagger}V_{\mathrm{F}}}{\sqrt{n^{0}}}=U,\label{eq:CF_U_equality}
\end{equation}

\noindent where we have used Eqs. (\ref{eq:Docc_rel}) and (\ref{eq:VF_rel}).

\subsection{Density matrix\label{subsec:Density-matrix}}

Using the block decomposition of the lattice density matrix:
\begin{equation}
\rho=\left[\begin{array}{cc}
\rho_{\mathrm{F}} & \rho_{c}\\
\rho_{c}^{\dagger} & \rho_{\mathrm{E}}
\end{array}\right]\label{eq:rho_blocks}
\end{equation}

\noindent (with $\rho_{\mathrm{F}}$ a $\mathcal{N}_c\times \mathcal{N}_c$ matrix, and so on)
and the expressions (\ref{eq:rho_diag}-\ref{eq:D_F_E}-\ref{eq:D_F_occ}),
we obtain:

\[
\rho=\left[\begin{array}{cc}
D_{\mathrm{F}}^{\mathrm{occ}}D_{\mathrm{F}}^{\mathrm{occ},\dagger} & D_{\mathrm{F}}^{\mathrm{occ}}D_{\mathrm{E}}^{\mathrm{occ},\dagger}\\
D_{\mathrm{E}}^{\mathrm{occ}}D_{\mathrm{F}}^{\mathrm{occ},\dagger} & D_{\mathrm{E}}^{\mathrm{occ}}D_{\mathrm{E}}^{\mathrm{occ},\dagger}
\end{array}\right].
\]

\noindent Thus, using Eq.~(\ref{eq:rho_embed_def}):
\begin{align}
 & \overline{\rho}_{\mathrm{embed}}\nonumber \\
 & =\left[\begin{array}{cc}
C_{\mathrm{F}}^{\dagger}D_{\mathrm{F}}^{\mathrm{occ}}D_{\mathrm{F}}^{\mathrm{occ},\dagger}C_{\mathrm{F}}, & C_{\mathrm{F}}^{\dagger}D_{\mathrm{F}}^{\mathrm{occ}}D_{\mathrm{E}}^{\mathrm{occ},\dagger}C_{\mathrm{B}}\\
C_{\mathrm{B}}^{\dagger}D_{\mathrm{E}}^{\mathrm{occ}}D_{\mathrm{F}}^{\mathrm{occ},\dagger}C_{\mathrm{F}}, & C_{B}^{\dagger}D_{\mathrm{E}}^{\mathrm{occ}}D_{\mathrm{E}}^{\mathrm{occ},\dagger}C_{\mathrm{B}}
\end{array}\right]\nonumber \\
 & =\left[\begin{array}{cc}
\frac{V_{\mathrm{F}}^{\dagger}D_{\mathrm{F}}^{\mathrm{occ,\dagger}}D_{\mathrm{F}}^{\mathrm{occ}}}{\sqrt{n^{0}}}\frac{D_{\mathrm{F}}^{\mathrm{occ},\dagger}D_{\mathrm{F}}^{\mathrm{occ}}V_{\mathrm{F}}}{\sqrt{n^{0}}}, & \frac{V_{\mathrm{F}}^{\dagger}D_{\mathrm{F}}^{\mathrm{occ,\dagger}}D_{\mathrm{F}}^{\mathrm{occ}}}{\sqrt{n^{0}}}\frac{D_{\mathrm{E}}^{\mathrm{occ},\dagger}D_{\mathrm{E}}^{\mathrm{occ}}V_{\mathrm{F}}}{\sqrt{n^{0}}}\\
\frac{V_{\mathrm{F}}^{\dagger}D_{\mathrm{E}}^{\mathrm{occ,\dagger}}D_{\mathrm{E}}^{\mathrm{occ}}}{\sqrt{1-n^{0}}}\frac{D_{\mathrm{F}}^{\mathrm{occ},\dagger}D_{\mathrm{F}}^{\mathrm{occ}}V_{\mathrm{F}}}{\sqrt{n^{0}}}, & \frac{V_{\mathrm{F}}^{\dagger}D_{\mathrm{E}}^{\mathrm{occ,\dagger}}D_{E}^{\mathrm{occ}}}{\sqrt{1-n^{0}}}\frac{D_{E}^{\mathrm{occ},\dagger}D_{\mathrm{E}}^{\mathrm{occ}}V_{\mathrm{F}}}{\sqrt{n^{0}}}
\end{array}\right]\nonumber \\
 & =\left[\begin{array}{cc}
n^{0} & \sqrt{n^{0}(1-n^{0})}\\
\sqrt{n^{0}(1-n^{0})} & 1-n^{0}
\end{array}\right].\label{eq:rho_embed_final}
\end{align}

\subsection{Parameters of the embedded Hamiltonian\label{subsec:Parameters-of-the}}

Identifying the blocks of Eqs~(\ref{eq:h_embed_projection}) and (\ref{eq:h_embed_parameters}),
and using Eq (\ref{eq:Pbar_def}), we obtain:
\begin{align}
\mathcal{D} & =h_{c}C_{\mathrm{B}}C_{\mathrm{F}}^{\dagger}=D_{\mathrm{F}}\epsilon D_{\mathrm{E}}^{\dagger}\frac{D_{\mathrm{E}}^{\mathrm{occ}}V_{\mathrm{F}}}{\sqrt{1-n^{0}}}\frac{V_{\mathrm{F}}^{\dagger}D_{\mathrm{F}}^{\mathrm{occ,\dagger}}}{\sqrt{n^{0}}},\label{eq:D_expr}\\
\lambda^{c} & =-C_{\mathrm{F}}C_{\mathrm{B}}^{\dagger}h_{\mathrm{E}}C_{\mathrm{B}}C_{\mathrm{F}}^{\dagger},\label{eq:lambda_expr}\\
 & =-\frac{D_{\mathrm{F}}^{\mathrm{occ}}V_{\mathrm{F}}}{\sqrt{n^{0}}}\frac{V_{\mathrm{F}}^{\dagger}D_{\mathrm{E}}^{\mathrm{occ,\dagger}}}{\sqrt{1-n^{0}}}D_{\mathrm{E}}\epsilon D_{\mathrm{E}}^{\dagger}\frac{D_{\mathrm{E}}^{\mathrm{occ}}V_{\mathrm{F}}}{\sqrt{1-n^{0}}}\frac{V_{\mathrm{F}}^{\dagger}D_{\mathrm{F}}^{\mathrm{occ,\dagger}}}{\sqrt{n^{0}}}.\nonumber 
\end{align}

\noindent Let us simplify $\mathcal{D}$:

\begin{align}
\mathcal{D} & =D_{\mathrm{F}}\epsilon D_{\mathrm{E}}^{\dagger}D_{\mathrm{E}}^{\mathrm{occ}}V_{\mathrm{F}}U^{\dagger}\left[\sqrt{(1-\Delta^{p})\Delta^{p}}\right]^{-1}UV_{\mathrm{F}}^{\dagger}V_{\mathrm{F}}\sqrt{n^{0}}U^{\dagger}\nonumber \\
 & =D_{\mathrm{F}}\epsilon D_{\mathrm{E}}^{\dagger}D_{\mathrm{E}}^{\mathrm{occ}}D_{\mathrm{F}}^{\mathrm{occ},\dagger}U\left[\sqrt{n^{0}}\right]^{-1}U^{\dagger}\left[\sqrt{(1-\Delta^{p})}\right]^{-1}\nonumber \\
 & =h_{c}\rho_{c}^{\dagger}\left[\sqrt{\Delta^{p}(1-\Delta^{p})}\right]^{-1}.\label{eq:D_expr-1}
\end{align}

\noindent Besides,
\begin{align*}
\left[h_{c}\rho_{c}^{\dagger}\right]_{\alpha\beta} & =\sum_{j\gamma}\left[t_{c}\right]_{\alpha,j\gamma}\left[\rho_{c}^{\dagger}\right]_{j\gamma,\beta}\\
 & =\sum_{j\gamma}\left[\tilde{t}\right]_{1\alpha,j\gamma}\left[\rho\right]_{j\gamma,1\beta}\\
 & =\sum_{\boldsymbol{k},\boldsymbol{k}',\gamma}\sum_{j}e^{i(\boldsymbol{k}-\boldsymbol{k}')\cdot\boldsymbol{R}_{j}}\left[\tilde{\varepsilon}(\boldsymbol{k})\right]_{\alpha,\gamma}\left[\rho(\boldsymbol{k}')\right]_{\gamma,\beta}\\
 & =\sum_{\boldsymbol{k},\gamma}\left[\tilde{\varepsilon}(\boldsymbol{k})\right]_{\alpha,\gamma}\left[\rho(\boldsymbol{k})\right]_{\gamma,\beta}\\
 & =\sum_{\boldsymbol{k},\gamma}\left[\tilde{\varepsilon}(\boldsymbol{k})\right]_{\alpha,\gamma}\sum_{i\omega}e^{i\omega0^{+}}\left[i\omega\boldsymbol{1}-\tilde{\varepsilon}(\boldsymbol{k})-\tilde{u}\right]_{\gamma,\beta}^{-1}.
\end{align*}

In the first line, we have used the block structure (\ref{eq:h_def}).
The second line follows from the definition of $t_{c}$ and $\tilde{t}$,
the third line from the definition of the Fourier transform, Eq. (\ref{eq:FT_def}),
and the fifth line from Eq. (\ref{eq:rho_noninteracting}). This yields
Eq. (\ref{eq:D_final_DMET}) of the main text.

\noindent Comparing with Eq. (\ref{eq:local_kinetic_energy_def}), we note that:
\begin{equation}
h_{c}\rho_{c}^{\dagger}=\mathcal{K}_{\mathrm{loc}}[R=\boldsymbol{1},\lambda=\tilde{u}].\label{eq:K_def}
\end{equation}

\noindent Let us now simplify $\lambda^{c}$: 

\begin{align*}
\lambda^{c} & =-U\sqrt{n^{0}}V_{\mathrm{F}}^{\dagger}V_{\mathrm{F}}U^{\dagger}\left[\sqrt{\Delta^{p}(1-\Delta^{p})\Delta^{p}}\right]^{-1}UV_{\mathrm{F}}^{\dagger}D_{\mathrm{E}}^{\mathrm{occ,\dagger}}\\
 & \;\times D_{\mathrm{E}}\epsilon D_{\mathrm{E}}^{\dagger}D_{\mathrm{E}}^{\mathrm{occ}}V_{\mathrm{F}}U^{\dagger}\left[\sqrt{(1-\Delta^{p})\Delta^{p}}\right]^{-1}UV_{\mathrm{F}}^{\dagger}V_{\mathrm{F}}\sqrt{n^{0}}U^{\dagger}\\
 & =-\left[\sqrt{(1-\Delta^{p})}\right]^{-1}U\left[\sqrt{n^{0}}\right]^{-1}U^{\dagger}D_{\mathrm{F}}^{\mathrm{occ}}D_{\mathrm{E}}^{\mathrm{occ,\dagger}}\\
 & \;\times D_{\mathrm{E}}\epsilon D_{\mathrm{E}}^{\dagger}D_{\mathrm{E}}^{\mathrm{occ}}D_{\mathrm{F}}^{\mathrm{occ},\dagger}U\left[\sqrt{n^{0}}\right]^{-1}U^{\dagger}\left[\sqrt{(1-\Delta^{p})}\right]^{-1}\\
 & =-\left[\sqrt{(1-\Delta^{p})\Delta^{p}}\right]^{-1}\rho_{c}h_{\mathrm{E}}\rho_{c}^{\dagger}\left[\sqrt{\Delta^{p}(1-\Delta^{p})}\right]^{-1}.
\end{align*}

\noindent To simplify $\rho_{c}h_{\mathrm{E}}\rho_{c}^{\dagger}$, let us first
notice:
\begin{align*}
\mathcal{K}_{\mathrm{loc}} & =h_{c}\rho_{c}^{\dagger}=D_{\mathrm{F}}\epsilon D_{\mathrm{E}}^{\dagger}D_{\mathrm{E}}nD_{\mathrm{F}}^{\dagger}\\
 & =D_{\mathrm{F}}\epsilon nD_{\mathrm{F}}^{\dagger}-D_{\mathrm{F}}\epsilon D_{\mathrm{F}}^{\dagger}D_{\mathrm{F}}nD_{\mathrm{F}}^{\dagger}\\
 & =D_{\mathrm{F}}\epsilon nD_{\mathrm{F}}^{\dagger}-h_{\mathrm{F}}\Delta^{p}.
\end{align*}

\noindent Hence:

\begin{align*}
\rho_{c}h_{\mathrm{E}}\rho_{c}^{\dagger} & =D_{\mathrm{F}}nD_{\mathrm{E}}^{\dagger}D_{\mathrm{E}}\epsilon D_{\mathrm{E}}^{\dagger}D_{\mathrm{E}}nD_{\mathrm{F}}^{\dagger}\\
 & =D_{\mathrm{F}}n\epsilon nD_{\mathrm{F}}^{\dagger}+D_{\mathrm{F}}nD_{\mathrm{F}}^{\dagger}D_{\mathrm{F}}\epsilon D_{\mathrm{F}}^{\dagger}D_{\mathrm{F}}nD_{\mathrm{F}}^{\dagger}\\
 & \;-D_{\mathrm{F}}n\epsilon D_{\mathrm{F}}^{\dagger}D_{\mathrm{F}}nD_{\mathrm{F}}^{\dagger}-D_{\mathrm{F}}nD_{\mathrm{F}}^{\dagger}D_{\mathrm{F}}\epsilon nD_{\mathrm{F}}^{\dagger}\\
 & =\mathcal{K}_{\mathrm{loc}}+h_{\mathrm{F}}\Delta^{p}+\Delta^{p}h_{\mathrm{F}}\Delta^{p}\\
 & \;-\left(\mathcal{K}_{\mathrm{loc}}+h_{\mathrm{F}}\Delta^{p}\right)\Delta^{p}-\Delta^{p}\left(\mathcal{K}_{\mathrm{loc}}+h_{\mathrm{F}}\Delta^{p}\right)\\
 & =\frac{1}{2}\left[\mathcal{K}_{\mathrm{loc}}(1-2\Delta^{p})+(1-2\Delta^{p})\mathcal{K}_{\mathrm{loc}}\right]\\&\;\;+\Delta^{p}(1-\Delta^{p})h_{\mathrm{F}}.
\end{align*}

\noindent This yields Eq. (\ref{eq:lambda_c_final}) of the main text.

\section{Ground-State energy and finite-temperature extension}\label{sec:gs_energy}

At $T=0$, the total energy in RISB is given by

\begin{eqnarray}
E&=&\sum_{\boldsymbol{k}}\mathrm{Tr}\left[n_\mathrm{F}\left(R\tilde{\varepsilon}_{\boldsymbol{k}}R^{\dagger}+\lambda-\mu\right)\left(R\tilde{\varepsilon}_{\boldsymbol{k}}R^{\dagger}\right)\right]\nonumber\\
&&+\sum_i \mathrm{Tr}\left[H_{\mathrm{loc}}\Phi_i\Phi_i^{\dagger}\right],\label{eq:GS_energy_RISB}
\end{eqnarray}

\noindent where $n_\mathrm{F}$ is the Fermi function, and $H_\mathrm{loc}$ contains the chemical potential $\mu$.

Note that it is straightforward to show that, using Eqs.~\eqref{eq:D_lagrange}, \eqref{eq:F2} and \eqref{eq:Trace_equivalence}, and with $R=1$, Eq.~\eqref{eq:GS_energy_RISB} is equivalent to the DMET ground-state energy given in~\cite{Knizia2012,Zheng2017}:

\begin{equation}
E=\sum_i\left\lbrace\mathrm{Tr}\sum_{\alpha\beta}\left(\mathcal{D}_{\alpha\beta}\Phi_i^{\dagger}F_{\alpha}^{\dagger}\Phi_i F_{\beta} + \mathrm{h.c.} \right)+\mathrm{Tr}\left[H_{\mathrm{loc}}\Phi_i\Phi_i^{\dagger}\right]\right\rbrace.\label{eq:GS_energy_DMET}
\end{equation}

In RISB, Eqs. \eqref{eq:GS_energy_RISB} and \eqref{eq:GS_energy_DMET} produce the same total energies because the Lagrange equations, Eqs. (\ref{eq:Delta_p-lagrange})- (\ref{eq:F2}), are exactly satisfied. However, in DMET, since the Lagrange equation, Eq. (\ref{eq:root_DMET}), can merely be minimized, 
Eqs. (\ref{eq:GS_energy_RISB}) and (\ref{eq:GS_energy_DMET}) no longer yield the same energy.
 One has to evaluate the total energy using Eq. (\ref{eq:GS_energy_DMET}) as done in the DMET literature.

The RISB formalism can be readily extended to finite temperatures, as will be explored in a separate publication. We give the final expression for the resulting free energy:

\begin{eqnarray}
\Omega&=&-T\sum_{\boldsymbol{k}}\log\left(1+e^{-\beta\left(R\tilde{\varepsilon}_{\boldsymbol{k}}R^{\dagger}+\lambda-\mu\right)}\right)\nonumber \\
&&+ T\sum_i \mathrm{Tr}\mathrm{log}\left[1-e^{-\beta( \hat{H}_{\mathrm{embed}}-E^c\hat{I})}\right]+E^c\nonumber\\
&&-\sum_{i,\alpha\beta} \left(\lambda_{\alpha\beta}+\lambda^c_{\alpha\beta}\right) \Delta^p_{\alpha\beta}\nonumber \\
&&-\sum_{i;\alpha\beta\gamma}\left(\mathcal{D}_{\alpha\beta}R_{\gamma\alpha}+\mathrm{c.c}\right)\left(\Delta^{p}(1-\Delta^{p})\right)_{\gamma\beta}^{1/2},\label{eq:free_energy_RISB}
\end{eqnarray}

\noindent where $\hat{I}$ is an identity matrix with the size of the Hilbert space of $\hat{H}_{\mathrm{embed}}$. 

The fact that DMET is a simplification of RISB with $R = 1$ gives an easy way to generalize DMET to finite temperatures. The implications of this finite-temperature extension of RISB and DMET will be explored in a separate publication.

\bibliographystyle{apsrev4-1}
\bibliography{extracted}

\end{document}